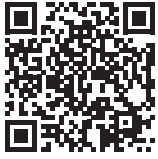

# Genetics and Pathophysiology of Maturity-onset Diabetes of the Young (MODY): A Review of Current Trends

Tajudeen O. Yahaya[1]* and Shemishere B. Ufuoma[2]

[1]Department of Biology, Federal University Birnin Kebbi, Kebbi State, Nigeria
[2]Department of Biochemistry and Molecular Biology, Federal University Birnin Kebbi, Kebbi State, Nigeria

ABSTRACT
Single gene mutations have been implicated in the pathogenesis of a form of diabetes mellitus (DM) known as the maturity-onset diabetes of the young (MODY). However, there are diverse opinions on the suspect genes and pathophysiology, necessitating the need to review and communicate the genes to raise public awareness. We used the Google search engine to retrieve relevant information from reputable sources such as PubMed and Google Scholar. We identified 14 classified MODY genes as well as three new and unclassified genes linked with MODY. These genes are fundamentally embedded in the beta cells, the most common of which are HNF1A, HNF4A, HNF1B, and GCK genes. Mutations in these genes cause β-cell dysfunction, resulting in decreased insulin production and hyperglycemia. MODY genes have distinct mechanisms of action and phenotypic presentations compared with type 1 and type 2 DM and other forms of DM. Healthcare professionals are therefore advised to formulate drugs and treatment based on the causal genes rather than the current generalized treatment for all types of DM. This will increase the effectiveness of diabetes drugs and treatment and reduce the burden of the disease.

Maturity-onset diabetes of the young (MODY) is a monogenic and non-autoimmune form of diabetes mellitus (DM) with characteristic pancreatic β-cell destruction and disrupted insulin biosynthesis.[1,2] The disease usually appears between the teen ages and early adulthood, < 25 years.[3,4] MODY was discovered by Robert Tattersall in 1974 as a distinct form of DM after discovering young non-insulin dependent diabetic individuals two-years post-diagnosis.[5,6] The condition was later named MODY by Fajans Stefan after a series of studies.[5,6] However, this classification may be confusing due to the similar pathophysiology of MODY and type 2 DM (T2DM). Many physicians and researchers often considered and misdiagnosed MODY as a subset of T2DM.[7] It was estimated that at least 90% of MODY diabetics are misdiagnosed as having T2DM due to lack of awareness on the differences between the two.[2,8,9] Some distinct characteristics of MODY include less significant weight gain, the absence of pancreatic autoantibodies, and lack of insulin resistance or elevated fasting glucose.[10] However, these symptoms are often unrecognized due to the low incidence of MODY, which is estimated to account for 1–5% of DM cases.[11–13]

MODY genes disrupt insulin production processes, culminating in hyperglycemia, which with time, may damage organs such as eyes, kidneys, nerves, and blood vessels.[14] The phenotypic expressions of MODY depend on the causal gene. Individuals with certain types of mutations may show a slightly raised blood sugar for life with mild or no symptoms of DM.[14] These individuals may not also develop long-term complications, and their high blood glucose levels may only be discovered during routine blood tests.[14] People with other mutation types require specific treatment with either insulin or a type of oral DM medication called sulfonylureas.[14] In the past, people with MODY had generally not been overweight or obese, or have other risk factors for T2DM, such as high blood pressure or abnormal blood fat levels.[14] However, as more people become overweight or obese, especially in the US, people with MODY may also be overweight or obese.[14] Although both T2DM and MODY can have a family history, such inheritance is autosomal dominant in MODY, meaning that, it does not skip any generation.[14]

*Corresponding author: yahaya.tajudeen@fubk.edu.ng

DM is a major public health problem worldwide.[15,16] The number of people with DM has quadrupled in the past three decades.[17] DM is now the ninth primary cause of death, accounting for about 1 in 11 adults with DM.[17] In 2019, about 463 million people were diagnosed with DM and this number was projected to reach 578 million by 2030, and 700 million by 2045.[18] More information and awareness are needed to contain the disease effectively. To this end, this study was conceived to provide current information on the already identified MODY genes with their mechanism of actions and phenotypic presentations. This will enable healthcare providers to formulate effective drugs and treatment methods for various forms of MODY.

### MODY genes
We identified 14 classified MODY genes as well as three new and unclassified genes linked with MODY. The identities of these genes with various mechanisms of action and phenotypic features are presented in Table 1.

### Common MODY genes
Using linkage analysis, restriction fragment length polymorphism, and DNA sequencing,[6] scientists have identified mutations in the hepatocyte nuclear factor 1-alpha (HNF1A), 4-alpha (HNF4A), 1-Beta (HNF1B), and glucokinase (GCK) genes as the most common cause of MODY.[48] Depending on the country, these genes account for over 80% of all MODY cases.[49–51]

### The HNF1A gene
The HNF1A gene provides instructions for the synthesis of the HNF1A protein.[52,53] The protein plays a vital role in the development of beta cells and the expression of many genes embedded in the liver.[52,53] These roles enable the pancreas to produce insulin normally in childhood, which decreases as one ages.[22] Thus, mutations in this gene may lower the amount of insulin produced,[22] and have been implicated in the pathogenesis of MODY type 3 (MODY3).[52,53] MODY3 is the commonest form of MODY, accounting for about 70% of cases.[52,53]

Several single nucleotide polymorphisms (SNPs) have been identified in the HNF1A gene of MODY patients, which could suggest the pathophysiology and possible treatment options. In a study, the coding and promoter regions of the HNF1A gene were screened for mutations in 34 unrelated Iranian patients with MODY. The study identified one novel missense mutation (C49G), two novel polymorphisms, and eight recently identified SNPs.[22] In another study, mutations identified in 356 unrelated MODY3 patients, including 118 novel mutations, were analyzed, and the correlation was drawn between the variants and age of onset of DM. Missense mutations were observed in 74% of cases, while 62% of patients had truncating mutations.[54] Most mutations (83%) were found in exons 1–6, wherein all three HNF1A isoforms are located and are thus affected.[54] The age of onset of DM was lower in patients with truncating mutations than in those with missense mutations.[54] It was also observed that the higher the number of HNF1A isoforms with missense mutations, the lower the age of diagnosis of DM in the patients.[54] These findings indicate that MODY3 patients may express variable clinical features depending on the type and location of the HNF1A mutations.[54] Aside from the liver and pancreas, HNF1A is embedded in the kidney, isolated islets, and intestines. So, the clinical presentations of individuals with HNF1A mutation may also depend on the tissues and its developmental stage.[55,56]

### The HNF4A gene
The HNF4A gene codes for a transcription protein embedded in the liver.[57,58] The HNF4A gene regulates the expression of several liver-specific genes. Thus, some liver functions may be enabled or disabled, depending on the expression or otherwise of this gene.[57,58] In addition, HNF4A controls the expression of the HNF1A gene, which in turn regulates the expression of several important genes in the liver.[57] The HNF4A gene is also found in the pancreas, kidneys, and intestines and, together with transcription factors such as HNF1A and HNF1B, control gene expression in developing embryos.[59,60] Specifically, in the pancreatic beta cells, this group of transcription factors controls the expression of the insulin gene. These genes also regulate the expression of several other genes involved in insulin secretion, such as the genes that are involved in glucose transport and metabolism.[59,61] Considering these functions, mutations in the HNF4A gene would lead to several problems.

**Table 1:** Maturity-onset diabetes of the young (MODY) genes showing chromosomal location and pathophysiology.

| Gene/function | Full name | Locus | MODY type | Pathophysiology |
|---|---|---|---|---|
| HNF4A/transcription factor | Hepatocyte nuclear factor-4 alpha | 20q12 | MODY 1 | Causes progressive beta-cell dysfunction, leading to macrosomia and hyperinsulinemic hypoglycemia.[8,19] |
| GCK/glycolytic enzyme | Glucokinase | 7p15 | MODY 2 | Disrupts glucose sensing, leading to hyperglycemia.[20,21] |
| HNFIA/transcription factor | Hepatocyte nuclear factor-1 alpha | 12q24.31 | MODY 3 | Causes gradual beta-cell dysfunction, leading to reduced insulin production and progressive hyperglycemia.[21,22] |
| IPFI/PDX1/transcription factor | Insulin promoter factor/Pancreatic duodenal homeobox | 13q27.92 | MODY 4 | Causes pancreatic agenesis, beta-cell developmental errors, and defective insulin secretion.[23,24] |
| HNFIB/transcription factor | Hepatocyte nuclear factor 1B | 17q12 | MODY 5 | Results in dysfunctional pancreatic embryonic development, the formation of kidney cyst, and suppresses cytokine signaling 3.[25,26] |
| NEURODI/transcription factor | Neurogenic differentiation 1 | 2q31.3 | MODY 6 | Impairs pancreatic morphogenesis and beta-cell differentiation.[27,28] |
| KLFII/transcription factor | Krüppel-like factor 11 | 2p25.1 | MODY 7 | Disrupts the activation of some insulin promoters. It also suppresses the expression of certain free radical scavengers such as catalase and superoxide dismutase, disrupting pancreatic beta-cell function.[29,30] |
| CELL/lipase | Carboxyl ester lipase | 9q34 | MODY 8 | Alters C-terminal sequencing. It can also disrupt exocrine and endocrine functioning of pancreas.[6,31] |
| PAX4/Transcription factor | Paired box 4 | 7q32.1 | MODY 9 | Truncates embryonic beta-cell development, inhibiting beta-cell differentiation.[32,33] |
| INS/Insulin synthesis | Insulin hormone | 11p15.5 | MODY 10 | Causes molecular defects in the β-cell and increases endoplasmic reticulum (ER) stress, resulting in the synthesis of structurally altered (pre)proinsulin molecules and low insulin biosynthesis.[34,35] |
| BLK/B-cell receptor signaling and development, stimula | B-lymphocyte kinase | 8p23.1 | MODY 11 | Suppresses MIN6 B-cells, disrupting beta-cell functions.[36,37] |
| ABCC8/regulates insulin secretion | ATP binding cassette subfamily C member 8 | 11p15.1 | MODY 12 | Causes congenital hyperinsulinism, adversely affecting the biogenesis and insulin trafficking of $K_{ATP}$ channels.[38,39] |
| KCNJII/regulates insulin secretion | Inward-rectifyier potassium channel, subfamily J, member 11 | 11p15.1 | MODY 13 | Causes congenital hyperinsulinism, adversely affecting the biogenesis and insulin trafficking of $K_{ATP}$ channels.[39,40] |
| APPL1/regulates cell proliferation, cellular signaling pathways | Adaptor protein, Phosphotyrosine interacting with PH domain and Leucine Zipper 1 | 3p14.3 | MODY 14 | Starts off the beta-cell structural abnormality and gradual death, leading to developmental delay. It can also suppress the insulin-uptake regulatory role of AKT2.[41,42] |
| ISL-1/transcription factor, INS enhancer | ISL LIM homeobox 1 | 5q11 | - | Interferes with the expression of several genes, including insulin gene, also causes poor islet differentiation and proliferation.[43,44] |
| RFX6/Regulatory factor (regulates the transcription factors involved in beta-cell maturation and function) | Regulatory factor X | 6q22.1 | - | Causes beta-cell dysfunction, leading to reduced insulin secretion and hyperglycemia.[45,46] |
| NK6-1/transcription factor | NK6 homeobox 1 | 4q21.23 | - | Beta-cell dysfunction.[47] |

Among the likely consequence of mutation in the HNF4A gene is the development of DM. The pancreatic beta-cell is sensitive to the population of HNF4A present, and certain HNF4A haplotypes are being linked with altered insulin secretion.[62] In particular, mutations in the gene are suspected in the pathogenesis of MODY type 1 (MODY1).[8] Individuals with MODY1 respond normally to insulin, but express an impaired response to secreting insulin in the presence of glucose.[8] If this condition remained unchecked, insulin secretion decreases, leading to DM.[8]

Several types of nonsense and missense mutations in HNF4A characterized by a shortfall in insulin secretion have been observed to cause MODY1.[62] Similarly, the variant of the HNF4A gene inherited may influence the function of beta cells, increasing or decreasing insulin secretion.[62] A British study identified a haplotype that was linked with reduced disease risk.[62] Individuals with the 'reduced-risk' haplotype were strongly associated with increased insulin secretion and lower fasting glucose levels.[63] These findings suggest that a certain HNF4A haplotype might have the ability for increased insulin secretion and protective effects on DM.[63] This protective variant was identified upstream of the HNF4A coding region in an alternative promoter called P2, which lies 46 kb upstream of the P1 promoter. Though the two promoters help in the transcription of HNF4A, they play different roles in different cells. Both P1 and P2 are embedded in the pancreas, but P2 is the main transcription site in the beta cells, and a mutation of P2 is a cause of MODY.[63]

How HNF4A mutations cause β-cell dysfunction or lipid profile disruption in MODY1 is not fully understood. However, based on its role in glucose transport and glycolysis as well as lipid metabolism,[64] loss of function of the gene could result in low triglyceride. This could end in less expression of some genes involved in glucose biosynthesis and metabolism.[64] Mice with a mutated HNF4A gene have been reported to show impaired glucose-stimulated insulin secretion and altered intracellular calcium response, characteristic of MODY1.[65] These observations were suggestive of loss of insulin regulatory function of $K_{ATP}$ channel in the pancreatic β-cells of the mutated mice.[65]

### *The HNF1B gene*
The HNF1B gene encodes a protein (called a transcription factor) that attaches to certain parts of DNA and modulates the expression of other genes.[66,67] The HNF1B protein is found in many organs and tissues, including the lungs, livers, intestines, pancreas, kidneys, reproductive system, and urinary tract.[66,67] Researchers suggest that the protein may be instrumental in the development of these body parts;[66] hence, its inactivation may initiate a lot of diseases. Notable among the diseases linked to mutations in the HNF1B gene is the MODY type 5 (MODY5).[66]

To prove the association between HNF1B and MODY5, a team of researchers compared pluripotent stem cell lines from MODY5 individuals with cells grown from unaffected family members and healthy controls.[25] In MODY5, children who inherited the mutation from one parent grew a malformed and small pancreas, indicating that they developed DM usually aged < 25 years.[25] The use of pluripotent stem cells allowed researchers to replicate human pancreas development in cell culture.[25] The scientists observed that HNF1B gene mutation disrupted the embryonic pancreas development of the cell cultures, leading to beta-cell dysfunction and impaired insulin biosynthesis.[25] It was also observed that mutations in this gene initiated DM independently of other DM genes.[25] However, the differentiating cells up-regulated other pancreatic development genes to compensate for the HNF1B inactivation.[25] These cellular events were observed in numerous MODY5 cell lines compared with healthy family members and non-related healthy controls.[25] The scientists were of the opinion that these findings and a greater understanding of beta-cell development and disruption could lead to improved DM treatments. This discovery once again shows the diabetic population could be stratified into subgroups and treated individually based on the mechanism of the causal gene rather than the current generalized treatment methods.[25]

As HNF1B is expressed in several tissues during embryonic development, diabetic conditions associated with HNF1B mutations can stem from extra-pancreatic abnormalities. The commonly afflicted organ is the kidney, usually affected by renal cysts, which precede diabetic conditions.[68–70] Renal dysplasia and renal tract malformations had also been reported to precede diabetic syndrome in individuals with HNF1B gene mutations.[68–70] Other precursors of MODY1 include hypomagnesemia, mild genital tract anomalies, abnormal liver morphology, and enzymes, especially alanine aminotransferase (ALT) and gamma-glutamyl transferase.[68–70]

### GCK gene

The GCK gene encodes the enzyme glucokinase, a member of the hexokinase family.[71] The gene plays a central role in carbohydrate metabolism in that it catalyzes the first reaction of the glycolytic pathway, the conversion of glucose to glucose 6-phosphate.[71] Glucokinase is expressed along with glucose transporter 2 (GLUT2) by the pancreatic β-cells and catalyzes the production of glucose, enabling it to act as a glucose sensor for the beta cells.[71,72] Compared with other hexokinase members, glucokinase has a high uninterruptible transport capacity for glucose.[71] Glucokinase works together with the GLUT2 receptor in the liver and beta-cell and enhances rapid insulin-independent entry and metabolism of glucose.[71] This allows the liver to act as a reservoir for circulating glucose as well as helps the glucose sensing mechanism of the beta cells.[71]

Mutations in the GCK gene have been demonstrated to cause abnormal glucose sensing, resulting in a raised threshold for initiation of glucose-stimulated insulin secretion. This ends in stable and mild hyperglycemia without any threat of DM complications.[73,74] This form of DM is known as GCK-MODY, otherwise known as MODY type 2 (MODY2). However, the clinical presentation of the MODY may vary based on the type of mutation. Heterozygous inactivating mutations cause mild fasting hyperglycemia (the hallmark of GCK-MODY), while the homozygous inactivating mutations cause a more severe condition, resembling permanent neonatal diabetes mellitus.[73,74] Other GCK mutations up-regulate insulin production, characterized by hyperinsulinemic hypoglycemia. In contrast to other forms of DM, hyperglycemia in MODY2 does not deteriorate with age.[75]

GCK expression is tissue-specific. In the liver, GCK synthesis is directly proportional to the concentration of insulin, which rises and falls with the nutritive state of the body.[76] On the other hand, glucagon, another liver hormone, suppresses GCK expression.[76] In the β-cells, GCK expression is relatively constant regardless of the body's food intake and, by extension, insulin levels.[76] Considering the roles of GCK in glucose metabolism and insulin release, GCK mutations are expected to cause both hyperglycemia and hypoglycemia.[77] Heterozygous mutations in the gene may result in a reduced phosphorylation rate in the liver, decreasing the concentration of glycogen synthesized, and disrupting postprandial glucose regulation.[78] In β-cells, loss of function of the gene will impair insulin secretion regulation.[79]

### Selection of individuals for MODY genetic testing

Due to the overlapping characteristics of various forms of DM, some of the discussed MODY pathophysiology might be observed in individuals with type 1 DM (T1DM) and T2DM, making the selection for genetic testing difficult. However, diabetic teens and young adults with a multi-generational history, as well as non-ketotic insulin-sensitive hyperglycemia and the absence of autoantibodies, should consider testing for MODY.[80,81] Additionally, middle-aged with autosomal dominant history and symptoms and signs of T2DM, but showing no obesity, insulin resistance and fatty liver, should also consider MODY testing.[81,82]

A MODY probability calculator developed by researchers at Exeter, UK, can also be used as a guide to select diabetics for genetic screening.[83] In the model, diabetics below 35 years old are scored based on their responses to eight questions. These questions include sex, age at diagnosis and referral, body mass index (BMI), treatment option taken, time insulin treatment begins, glycated hemoglobin ($HbA_{1c}$) level, and diabetic status of the parents.[83,84] For MODY to be considered against T1DM, $HbA_{1c}$ must be lower, at least one of the parents must be affected, and the age of diagnosis must be older.[83,84] MODY will be considered against T2DM if the BMI is lower, the age of diagnosis is younger, $HbA_{1c}$ is lower, and if the affected has a diabetic parent(s).[83,84] MODY will also be suspected against T2DM if the diabetic does not respond to oral hypoglycemic drugs or insulin.[83,84]

In advanced countries, some organizations have compiled major clinical features of MODY and employed them as guidelines for the selection of diabetics for MODY testing. Notable among these organizations/guidelines include European Molecular Genetics Quality Network Best Practice Guidelines[82] and the Clinical Practice Consensus

Guidelines by the International Society for Pediatric and Adolescent Diabetes.[85]

However, comparing all these selection methods with the large numbers of MODY pathophysiology described in this study shows the methods lack adequate information to detect MODY accurately. This suggests that many MODY patients are still currently being misdiagnosed as T1DM or T2DM. Thus, the MODY population worldwide could be higher than those reported in several studies.

### Genetic testing techniques for MODY

When there are means and sufficient information that an individual has MODY, the next step is to choose a screening procedure or technique. For effective prevention and management of genetic diseases, including MODY, genetic testing should begin during the intrauterine life, which is termed prenatal genetic testing.[86] Prenatal testing can be carried out to detect genetic errors related to MODY during fetal development. Genetic testing for GCK-MODY is particularly important during pregnancy to confirm the presence or otherwise of macrosomia, which may help in the choice of therapies.[87,88] Genetic testing can also be done immediately after birth to detect MODY mutations that can be corrected if detected early. This form of genetic testing is called newborn testing.[86] Predictive and pre-symptomatic testing is also conducted a few years after birth[86] to identify the MODY risk of an individual, especially those with a family history of MODY. Diagnostic testing is another test that can be conducted at any time when certain biomarkers or pathophysiology of MODY are observed in an individual. The test is often used to confirm the status of a specific genetic or chromosomal condition.[86]

A couple of biological techniques are available for genetic testing, notable among which are gene-targeted testing (serial single gene or multigene panel) and whole-exome sequencing.[89]

### Gene-targeted testing (serial single gene or multigene panel)

Gene-targeted testing, such as Sanger sequencing, is genetic testing in which specific genes, based on the clinical presentations of the person with diabetes, are selected for testing.[90] The technique is most suitable when the patient expresses some signs related to a few known MODY types. The test is carried out serially in which a sequence analysis of the probable genes is carried out first.[89] For a patient showing the classical features of MODY, HNF1A is screened first, followed by HNF4A and GCK. However, if the diabetic phenotype is mild and fasting glucose is between 5.5 and 8.5 mmol/L, GCK should be tested first, then HNF1A and HNF4A in that order.[12,91] If the patient has renal and pancreatic dysfunction as well as a urogenital problem, HNF1B should be tested first.[12,91] If no SNP was discovered, deletion/duplication analysis should be done to detect such genes as CEL, GCK, HNF1A, HNF1B, and HNF4A.[89] Generally, the gene-targeted approach is time-consuming and expensive. Complete genetic testing for HNF1A, HNF4A, and GCK involves sequencing 31 exons in which each gene is sequenced separately.[92]

Alternatively, a MODY multigene panel that contains the 14 known MODY genes and other suspect genes can be employed to detect the genetic cause of a MODY. This is cost-effective as it targets several genes at a time and precludes testing unnecessary variants.[90,93]

### Whole-exome capture and high-throughput sequencing

When a MODY patient does not show sufficient or clear clinical features, in-depth genomic screening such as whole-exome sequencing could be the best genetic testing option.[89] In exome sequencing, the selection of probable genes for testing is not needed. Because of these reasons, exome sequencing has some advantages over gene-targeted sequencing in that it can detect MODY genes beyond the reach of the latter.[92] In some cases, exome sequencing is used as a further diagnostic tool where gene-targeted sequencing is ineffective due to insufficient clinical expressions it needs to work with. Whole-exome sequencing is relatively new and could be improved upon in the near future to expand its search coverage and potential.[92] If this is done, it will make MODY diagnosis easier and better.

### Cost-effectiveness of genetic testing for MODY

For individuals, the huge cost of genetic testing for MODY could be burdensome. However, if done accurately, it will improve the quality of life. Testing for MODY genes in a family with the disease may help detect MODY variants in predisposed members and proffer treatment before it degenerates to glucose imbalance and DM. Accurate genetic screening may help predict the likelihood and types of complications and, in turn, reduce expenditures. For instance, HNF1A and HNF4A MODY are characterized by microvascular complications, which can be managed with a low dose of sulfonylureas instead of rigorous insulin therapy[94] typified by T1DM. On the other hand, GCK-MODY shows less microvascular complications and may not need any treatment.[92] So, accurate diagnosis of MODY type could prevent a wrong treatment choice, culminating in reduced healthcare cost.[95,96]

In a society where there is insurance cover or policy for testing, the cost-effectiveness of genetic testing for MODY depends on the frequency of the condition in the population.[97,98] In a stimulated study carried out in the US by Naylor et al,[98] genetic testing for MODY was non-cost-effective when the frequency of the disease was as low as 2%. However, when the population of MODY was increased to 6% with improved screening techniques and expanded pathophysiology, testing was found to be cost-effective.[98] Moreover, genetic testing was found to be cost-effective in the population, with a 2% prevalence of MODY when the cost of testing was reduced.[98] The study also demonstrated that if the MODY population is increased to 31%, through advanced testing techniques, the genetic testing policy for MODY becomes cost-saving.[98] In brief, the cost-effectiveness of a genetic testing policy depends on the frequency of MODY in society and the cost of the test.

## CONCLUSION

An autosomal dominant mutation in certain genes involved in insulin biosynthesis and metabolism may cause MODY. This form of DM has distinct pathogenic and clinical presentations. Thus, its treatment may require a different approach from other types of disease. As such, healthcare providers are advised to formulate MODY drugs and treatment methods based on the identified mechanism of action and phenotypic presentations of its subtypes.

*Disclosure*
The authors declared no conflicts of interest. No funding was received for this study.